\documentstyle[12pt,epsfig]{article}
\newcommand{\dd}{\mbox{d}}
\newcommand{\eps}{\varepsilon}
\newcommand{\Li}{\mbox{Li}_2}
\newcommand{\vecc}[1]{\mbox{\boldmath $#1$}}
\def\fun#1#2{\lower3.6pt\vbox{\baselineskip0pt\lineskip.9pt
\ialign{$\mathsurround=0pt#1\hfil##\hfil$\crcr#2\crcr\sim\crcr}}}
\title{Radiative corrections for pion and kaon production
at $e^+e^-$ colliders of energies below 2 GeV}

\author{A.B.~Arbuzov$^{1}$, V.A.~Astakhov$^{2}$, A.V.~Fedorov$^{3}$
G.V.~Fedotovich$^{2}$, \\ E.A.~Kuraev$^{1}$ and N.P.~Merenkov$^{4}$}

\date{}

\begin{document}

\maketitle

\begin{center}
{$^1$ \it Bogoliubov Laboratory of Theoretical Physics, JINR, \\
Dubna, 141980, Russia} \\[.2cm]
{$^2$ \it Budker Institute for Nuclear Physics, \\
Prospect Nauki, 11, Novosibirsk, 630090, Russia}\\[.2cm]
{$^3$ \it Laboratory of Computing Techniques
and Automation, \\ JINR, Dubna, 141980, Russia}\\[.2cm]
{$^4$ \it Kharkov Institute of Physics and Technology, \\
Kharkov, 310108, Ukraine} \\[1.2cm]
\end{center}

\begin{abstract}
Processes of electron--positron annihilation into charged
pions and kaons are considered. Radiative corrections are
taken into account exactly in the first order and within
the leading logarithmic approximation in higher orders.
A combined approach for accounting exact calculations and
electron structure functions is used.
An accuracy of the calculation can be estimated about 0.2\%.
\\[.2cm] \noindent
PACS~ 12.20.--m Quantum electrodynamics, 12.20.Ds Specific calculations \\
%\begin{keyword}
\end{abstract}

%\end{frontmatter}

\section{Introduction}

The processes in collisions of $e^+e^-$ beams at moderate high energies
with detection of the final particles moving in centre--of--mass
system (c.m.s.) at large angles are the subject of close
attention at meson factories, such as VEPP--2M (Novosibirsk)~\cite{vepp2m},
DA$\Phi$NE (Frascati)~\cite{dafne}, BEPC/BES (Bejing)~\cite{bepc}.
The processes of
pure quantum electrodynamics (QED) nature provide an important
background for studies of subtle mesons properties. Besides,
they may be used for a calibration and monitoring.
Because of large cross--sections of the lowest order processes,
radiative corrections (RC) to them are to be included in the
consideration.

In our previous paper \cite{lqed} we had considered the QED processes
$e^+e^-\rightarrow e^+e^-(\gamma)$, $\mu^+\mu^-(\gamma)$,
$\gamma\gamma(\gamma)$. We developed there an approach
for precise accounting of radiative corrections to
differential distributions. The part of RC, describing the
emission of hard additional photons was presented in the form
convenient for imposing experimental conditions of the final particle
detection. The contribution of higher orders of perturbation theory
was considered in the leading logarithmical approximation.
That was done by means of the structure function approach,
writing a cross--section in the Drell--Yan form.

In this paper we apply the same approach to processes $e^+e^-
\rightarrow \pi^+\pi^-(\gamma)$, $K_LK_S(\gamma)$, $K^+K^-(\gamma)$,
considering the pseudoscalar mesons as point--like objects.
The effects of strong
interaction of hadrons in the final state are parametrized by introducing
form factors which are to be measured in an experiment.
We assume as usually that vacuum polarization corrections
(by hadrons and leptons) are also included in the form factors.

The virtual and soft real photon emission corrections are calculated
in the ${\cal O}(\alpha)$ order exactly. This permits us to keep
explicitly the leading (containing {\em large\/} logarithm
$L=\ln(s/m_e^2)$, $\ s=4\varepsilon^2$
is the squared total energy in c.m.s.) and
next--to--leading terms. The latters are regarded further
as ${\cal K}$--factor terms in the Drell--Yan representation
for a cross--section.

Considering hard photon emission we extract contributions,
containing large logarithm $L$. We distinguish the collinear kinematics of
hard photon emission and the semi--collinear ones, which do not
give rise to $L$. An auxiliary parameter, a small polar angle
$\theta_0\ll 1$ with respect to the direction of the initial beams,
is introduced for this purpose. The terms containing photon softness
parameter $\Delta=\Delta\varepsilon/\varepsilon$
($\Delta\varepsilon$ is the maximal energy of a soft (in c.m.s.) undetected
photon) and the ones containing $\theta_0$ from the contribution of the
collinear region are regarded as compensating terms. When they being summed
with the contribution of the semi--collinear region (with the relevant
restrictions imposed) provide their finiteness in the limit $\Delta,
\theta_0 \rightarrow 0$.

This paper is organized as follows. In the second Section
we consider the process of charged pion production.
The explicit formulae (we keep pion mass exactly)
for the lowest order virtual and soft real photon emission
are presented. The charge--even and charge--odd contributions
are given separately. The latter quantity permits us to
obtain the charge asymmetry which can be measured.
We put also the explicit formula for the differential cross--section
of hard photon emission.
In Sect.~3 a similar consideration is given for the case
of neutral and charged kaon production near the threshold.
In Conclusions we discuss the formulae obtained and their precision.
The results are illustrated numerically in Figures.
In the Appendix we discuss the structure of form factors and give the
explicit expressions for the vacuum polarization operator.

\section{Pion Pair Production}

In the Born approximation the differential cross--section
of the process
\begin{eqnarray}
e^+(p_+)\ +\ e^-(p_-)\ \to\ \pi^+(q_+)\ +\ \pi^-(q_-)
\end{eqnarray}
has the form
\begin{eqnarray}
&& \frac{\dd\sigma_0}{\dd\Omega}(s)=\frac{\alpha^2\beta^3}{8s}\sin^2\theta\;
|F_{\pi}(s)|^2, \qquad
\beta=\sqrt{1-m_{\pi}^2/\eps^2},\\ \nonumber
&& s=(p_++p_-)^2=4\eps^2, \qquad
\theta=\widehat{\vecc{p}_-\vecc{q}}_-\, .
\end{eqnarray}
The pion form factor $F_{\pi}(s)$ takes into account
vertex virtual corrections due to strong interactions
and vacuum polarization by leptons and hadrons
(including vector--meson resonances)~\cite{fpi} (see Appendix).
We would like to underline that in our approach QED
corrections are not included into $F_{\pi}(s)$.  One has to
take the form factor from an experiment after an extraction
of QED radiative corrections.

Calculating QED radiative corrections we will consider pion as
a point--like particle. We distinguish form--factor--type
one--loop Feynman diagrams and the box--type ones.
QED form factors are taken as
$F_{e,\pi}^{QED}(s)=1+F_{e,\pi}^{(1)}+{\cal O}(\alpha^2)$.
Using the known first order contributions
to the electron and pion QED form factors
\begin{eqnarray}
\mbox{Re}\;F_e^{(1)}(s)&=&\frac{\alpha}{\pi}
\biggl[\left(\ln\frac{m_e}{\lambda}-1\right)(1-L )
-\frac{1}{4}L^2+\frac{\pi^2}{3}-\frac{1}{4}L \biggr],\qquad
L = \ln\frac{s}{m_e^2}\, ,
\\ \nonumber
\mbox{Re}\;F^{(1)}_{\pi}(s) &=& 1 + \frac{\alpha}{4\pi}\biggl\{4
\left(\ln\frac{m_{\pi}}{\lambda}-1\right)\left(1
-\frac{1+\beta^2}{2\beta}\; l_{\beta}\right) \\ \nonumber
&+& \frac{1+\beta^2}{2\beta}\biggl[-2\ln\left(\frac{4}{1-\beta^2}\right)\;
l_{\beta} + \ln^2\frac{1-\beta}{2} - \ln^2\frac{1+\beta}{2} \\
&-& 2l_{\beta}\ln\beta - 2\mbox{Li}_2\left(-\frac{1-\beta}{2\beta}\right)
+ 2\mbox{Li}_2\left(\frac{1-\beta}{2\beta}\right) \biggr] \biggr\}
\end{eqnarray}
and the contribution due to soft photon emission~\cite{prep,int},
we obtain for the charge--even part of the differential cross--section
the following formula:
\begin{eqnarray} \label{evebsv}
&& \frac{\dd\sigma^{B+S+V}_{\mathrm{even}}}{\dd\Omega} =
\frac{\dd\sigma_0}{\dd\Omega}\biggl\{1+\frac{2\alpha}{\pi}[A+B]\biggr\},
\\ \nonumber
&& A = (L -1)\ln\frac{\Delta\eps}{\eps} + \frac{3}{4}(L  - 1 ) + a,
\qquad a=\frac{\pi^2}{6} - \frac{1}{4}, \\ \nonumber
&& B = \left(\frac{1+\beta^2}{2\beta}\ln\frac{1+\beta}{1-\beta}-1 \right)
\ln\frac{\Delta\eps}{\eps} + b(s), \\ \nonumber
&& b(s) = - 1 + \frac{1-\beta}{2\beta}L
+ \frac{1}{\beta}\ln\frac{1+\beta}{2} + \frac{1+\beta^2}{2\beta}
\biggl[-\mbox{Li}_2\left(-\frac{1-\beta}{1+\beta}\right) \\ \nonumber
&&\qquad + \mbox{Li}_2\left(\frac{1-\beta}{1+\beta}\right)
- \frac{\pi^2}{12}
+ L\ln\frac{1+\beta}{2} - 2L\ln\beta
+ \frac{3}{2}\ln^2\frac{1+\beta}{2}
\\ \nonumber
&&\qquad - \frac{1}{2}\ln^2\beta - 3\ln\beta\ln\frac{1+\beta}{2} + L
+ 2\ln\frac{1+\beta}{2}\biggr],
\qquad \Li(x)=-\int\limits_{0}^{x}\frac{\dd t}{t}\ln(1-t).
\end{eqnarray}

Box--type diagrams and the interference of soft photon emission
from electrons and pions give rise for the charge--odd contribution
\begin{eqnarray} \label{oddsv}
\frac{\dd\sigma^{S+V}_{\mathrm{odd}}}{\dd\Omega} &=&
\frac{\dd\sigma_0}{\dd\Omega}\;\frac{2\alpha}{\pi}\biggl\{
2\ln\frac{\Delta\eps}{\eps}\ln\frac{1-\beta c}{1+\beta c} + k(c,s)
\biggr\}, \\ \nonumber
k(c,s) &=& \frac{1}{2}l_-^2
- \mbox{Li}_2\left(\frac{1-2\beta c+\beta^2}{2(1-\beta c)}\right)
+ \mbox{Li}_2\left(\frac{\beta^2(1-c^2)}{1-2\beta c+\beta^2}\right)
\\ \nonumber
&-& \int\limits_{0}^{1-\beta^2}\frac{\dd x}{x}f(x)\left(1
- \frac{x(1-2\beta c+\beta^2)}{(1-\beta c)^2}\right)^{-\frac{1}{2}}
\\ \nonumber
&+& \frac{1}{2\beta^2(1-c^2)}\Biggl\{\left[\frac{1}{2}l_-^2
- (L+l_-)L_-
+ \Li\left(\frac{1-\beta^2}{2(1-\beta c)}\right) \right]
(1-\beta^2) \\ \nonumber
&+& (1-\beta c)\biggl[-l_-^2
- 2\Li\left(\frac{1-\beta^2}{2(1-\beta c)}\right)
+ 2(L+l_-)L_-
- \frac{(1-\beta)^2}{2\beta}\left(\frac{1}{2}L^2
+ \frac{\pi^2}{6}\right) \\ \nonumber
&+& \frac{1+\beta^2}{\beta}\biggl(
L\ln\frac{2}{1+\beta}
- \Li\left(-\frac{1-\beta}{1+\beta}\right)
+ 2\Li\left(\frac{1-\beta}{2}\right) \biggr] \Biggr\}
- (c\to -c), \\ \nonumber
f(x)&=&\biggl(\frac{1}{\sqrt{1-x}}-1\biggr)
\ln\frac{\sqrt{x}}{2} - \frac{1}{\sqrt{1-x}}\ln\frac{1+\sqrt{1-x}}{2}\, ,
\\ \nonumber
l_-&=&\ln\frac{1-\beta c}{2}\, ,\qquad
L_-=\ln\biggl(1-\frac{1-\beta^2}{2(1-\beta c)}\biggr).
\end{eqnarray}
The charge asymmetry in a quasi--elastic case has the form
\begin{equation}
\eta=\frac{\dd\sigma(c)-\dd\sigma(-c)}{\dd\sigma(c)+\dd\sigma(-c)}
= \frac{\dd\sigma^{S+V}_{\mathrm{odd}}}{\dd\sigma_{0}}\, .
\end{equation}
In the ultra--relativistic case $(\beta\to 1)$ we obtain
\begin{eqnarray}
(\eta)_{\mbox{\scriptsize asympt}}&=&\frac{2\alpha}{\pi}
\biggl[4\ln (\mbox{tg}\frac{\theta}{2})\,
\ln \frac{\Delta\eps}{\eps}+\left(2-\frac{1}{\cos^2\frac{\theta}{2}}\right)
\ln^2(\sin\frac{\theta}{2}) \\ \nonumber
&-& \biggl(2-\frac{1}{\sin^2\frac{\theta}{2}}\biggr)
\ln^2(\cos\frac{\theta}{2})+\mbox{Li}_2(\cos^2\frac{\theta}{2})-
\mbox{Li}_2(\sin^2\frac{\theta}{2})\biggr],\quad \eps\gg m.
\end{eqnarray}
This expression coincides with the result of Brown and Mikaelian~\cite{brown}.

The matrix element of the process accompanied by hard photon
emission
\begin{eqnarray}
e^-(p_-)+e^+(p_+)\longrightarrow \pi^-(q_-) + \pi^+(q_+) + \gamma(k)
\end{eqnarray}
can be presented in the following form:
\begin{eqnarray}
&& M^{e^+e^-\to\pi^+\pi^-\gamma} = -i(4\pi\alpha)^{\frac{3}{2}}
\Biggl\{ \bar{v}\biggl[\gamma_{\nu}\left(\frac{p_+e}{\chi_+}
- \frac{p_-e}{\chi_-}\right)
+ \frac{\gamma_\nu\hat{k}\hat{e}}{2\chi_-}
- \frac{\hat{e}\hat{k}\gamma_\nu}{2\chi_+} \biggr] \\ \nonumber
&&\qquad \times u(q_- -q_+)^{\nu} \frac{F_{\pi}(s_1)}{s_1}
+ \bar{v}\gamma_{\rho}u \frac{F_{\pi}(s)}{s}\, T^{\pi}_{\rho\sigma}
e^{\sigma}(k) \Biggr\}, \\ \nonumber
&& p_-+ p_+ = q_- + q_+ + k,\quad s = (p_-+p_+)^2,\quad s_1 = (q_-+q_+)^2,
\quad \chi_{\pm} = p_{\pm}k.
\end{eqnarray}
Tensor $T^{\pi}_{\rho\sigma}$ describes the transition of a {\em heavy\/}
photon into the system of two real pions and a real photon:
\begin{eqnarray}
&& \gamma^*(q) \rightarrow \pi^+(q_+) + \pi^-(q_-) + \gamma(k), \quad
q_+^2=q_-^2=m_{\pi}^2, \quad q^2=s, \quad k^2=0.
\end{eqnarray}

Regarding $CPT$ and gauge invariance the tensor can be written
in the general form
\begin{eqnarray}
T^{\pi}_{\rho\sigma}=a_1L^{(1)}_{\rho\sigma} + a_2L^{(1)}_{\rho\sigma} +
a_3L^{(1)}_{\rho\sigma}+ k_{\sigma}O_{\rho}, \qquad
q^{\rho}T^{\pi}_{\rho\sigma}=0,\quad k^{\sigma}T^{\pi}_{\rho\sigma}=0.
\end{eqnarray}
The last term $(\sim k_{\sigma})$ is irrelevant here. Tensors $L^{(i)}$
read
\begin{eqnarray}
L^{(1)}_{\rho\sigma}&=&qk\; g_{\rho\sigma} - k_{\rho}q_{\sigma}, \qquad
L^{(2)}_{\rho\sigma}=qk\; Q_{\rho}Q_{\sigma}
- kQ\;(q_{\sigma}Q_{\rho}+Q_{\sigma}k_{\rho})+(kQ)^2g_{\rho\sigma},
\nonumber \\
L^{(3)}_{\rho\sigma}&=&kQ\;(q^2 g_{\rho\sigma}-q_{\rho}q_{\sigma})
+ Q_{\sigma}(qk\;q_{\rho} - q^2k_{\rho}), \qquad Q=\frac{1}{2}(q_1-q_2).
\end{eqnarray}
For the case of charge pion production, which is considered below,
we used the approximation of point--like pions, where
\begin{eqnarray}
T^{\pi}_{\rho\sigma}\to T^{(0)}_{\rho\sigma} &=&
\frac{1}{2\chi_+'}(q_--q_+-k)_{\rho}(-2q_+-k)_{\sigma} \nonumber \\
&+& \frac{1}{2\chi_-'}(q_--q_++k)_{\rho}(2q_-+k)_{\sigma}
- 2g_{\rho\sigma}\, ,\qquad \chi_{\pm}'=q_{\pm}k, \\ \nonumber
a_1^{(0)}&=&- \frac{2}{\chi_-'+\chi_+'}
\left(\frac{\chi_-'}{\chi_+'} + \frac{\chi_+'}{\chi_-'}\right),
\qquad a_2^{(0)}=\frac{16}{\chi_-'\chi_+'}\, ,
\qquad  a_3^{(0)}=0.
\end{eqnarray}
In reality some vector--meson resonance intermediate states
give rise of contributions to the tensor.

The differential cross--section of the process with hard photon
emission reads
\begin{eqnarray} \label{r123}
&& \dd\sigma^{e^+e^-\to\pi^+\pi^-\gamma}=
\frac{\alpha^3}{2\pi^2s^2}(R_1+R_{2}+R_{3})\dd\Gamma, \\ \nonumber
&& R_1=\frac{s}{s_1^2} |F_{\pi}(s_1)|^2
(p_+Qp_-Q)\biggl[\frac{p_+p_-}{\chi_+\chi_-}
- \frac{2}{\chi_-}-\frac{m_e^2}{\chi_-^2} \\ \nonumber && \qquad
+ \frac{\chi_+}{p_+p_-}\left(
\frac{1}{\chi_-} + \frac{m_e^2}{\chi_-^2}\right)
+ (p_+\leftrightarrow p_-) \biggr], \\ \nonumber
&& R_{2}=\biggl\{\frac{1}{s}|F_{\pi}(s)|^2\biggl[
\frac{q_+q_-}{\chi_+'\chi_-'} - \frac{m_{\pi}^2}{(\chi_+')^2}
+ (q_+ \leftrightarrow q_-) \biggr]  \\ \nonumber
&& \qquad + \frac{2}{s_1}\mbox{Re}\;(F_{\pi}(s)F^{*}_{\pi}(s_1))
\left(\frac{p_+}{\chi_+}-\frac{p_-}{\chi_-}\right)
\left(\frac{q_+}{\chi_+'}-\frac{q_-}{\chi_-'}\right)\biggr\}
(p_+Qp_-Q), \\ \nonumber
&& R_{3}=\frac{s}{s_1^2}|F_{\pi}(s_1)|^2\biggl[\frac{(p_+Qk_-Q)}{\chi_-}
+ \frac{(p_-Qk_+Q)}{\chi_+} + \frac{2Qk}{\chi_+\chi_-}(p_+Qp_-k)\biggr]
\\ \nonumber && \qquad
+ \frac{1}{s}|F_{\pi}(s)|^2\biggl[
- \frac{\chi_+\chi_-}{2\chi_+'\chi_-'}q_+q_-
- \frac{m_{\pi}^2}{4(\chi_+')^2}\biggl( (p_+kp_-k) + 2(p_+Qp_-k)
\\ \nonumber && \qquad
+ 2(p_+kp_-Q) \biggr)
+ \frac{1}{\chi_+'}\biggl(\chi_+p_-q_+ + \chi_-p_+q_+
+ 2(p_-Qp_+q_+) \biggr) + (q_+\leftrightarrow q_-) \biggr]
\\ \nonumber && \qquad
+ \frac{1}{s}\mbox{Re}\;(F_{\pi}(s)F^{*}_{\pi}(s_1))
\biggl\{(p_+Qp_-k)\frac{q_+}{\chi_+'}\left(
\frac{p_+}{\chi_+} - \frac{p_-}{\chi_-}\right) \\ \nonumber && \qquad
+ 2(p_+Q-p_-Q)
- \frac{1}{\chi_+'}\left((p_+Qq_+k)-(p_-Qq_+k)\right) \\ \nonumber && \qquad
+ \frac{2Qk\; Qp_+\; p_-q_+}{\chi_+'\chi_-}
- \frac{2Qp_-\; Qk\; p_+q_+}{\chi_+'\chi_+}
- \frac{Q^2\;\chi_+\; q_+p_-}{\chi_+'\chi_-}
\\ \nonumber && \qquad
+ \frac{Q^2\; p_+q_-}{\chi_+'}
+ \frac{Q^2\;\chi_-\; p_+q_+}{\chi_+'\chi_+}
- \frac{Q^2\; p_-q_-}{\chi_+'}
- (q_+ \leftrightarrow q_-) \biggr\}, \\ \nonumber
&& k_{\pm}=k-p_{\pm}\;\frac{\chi_{\mp}}{p_+p_-}\, ,\qquad
Q=\frac{1}{2}(q_+-q_-)\, , \\ \nonumber
&& t = -2p_-q_-,\quad t_1 = -2p_+q_+,\quad u = -2p_-q_+,
\quad u_1 = -2p_+q_-, \\ \nonumber
&& \dd\Gamma = \frac{\dd^3 q_+\dd^3 q_-\dd^3k}{q_-^0q_+^0k^0}
\delta^{(4)}(p_-+p_+-q_--q_+-k),
\end{eqnarray}
where we use the notation:
\begin{eqnarray}
(abcd)=\frac{1}{4}\mbox{Sp}\;\hat{a}\hat{b}\hat{c}\hat{d}=(ab)\;(cd)
+ (ad)\;(bc) - (ac)\;(bd).
\end{eqnarray}

After algebraic transformations one can get a more compact expression:
\begin{eqnarray} \label{rss}
&& \dd\sigma^{e^+e^-\to\pi^+\pi^-\gamma}=
\frac{\alpha^3}{32\pi^2s}(R_{s_1s_1}+R_{ss}+R_{ss_1})\;\dd\Gamma, \\ \nonumber
&& R_{s_1s_1} = |F_{\pi}(s_1)|^2\Biggl\{ A\,\frac{4s}{\chi_-\chi_+}
- \frac{8m_{e}^2}{s_1^2}\left(\frac{t_1u_1}{\chi_-^2}
+ \frac{tu}{\chi_+^2}\right) + m_{\pi}^2\Delta_{s_1s_1}\Biggr\}, \\ \nonumber
&& R_{ss}  = |F_{\pi}(s)|^2\Biggl\{ A\,\frac{4s_1}{\chi_-'\chi_+'}
- \frac{8m_{\pi}^2}{s^2}\left(\frac{tu_1}{(\chi_+')^2}
+ \frac{t_1u}{(\chi_-')^2}\right) + m_{\pi}^2\Delta_{ss}\Biggr\}, \\ \nonumber
&& R_{ss_1} = \mbox{Re}\;(F_{\pi}(s)F_{\pi}^*(s_1)) \Biggl\{
4A\left(\frac{u}{\chi_-\chi_+'}+\frac{u_1}{\chi_+\chi_-'}
- \frac{t}{\chi_-\chi_-'} - \frac{t_1}{\chi_+\chi_+'} \right)
+ m_{\pi}^2\Delta_{ss_1} \Biggr\}, \\ \nonumber
&& A = \frac{tu+t_1u_1}{ss_1}\, , \qquad
\Delta_{s_1s_1} = - \frac{4}{s_1^2}\,
\frac{(t+u)^2+(t_1+u_1)^2}{\chi_+\chi_-}\, , \\ \nonumber
&& \Delta_{ss} = \frac{2m_{\pi}^2(s-s_1)^2}{s(\chi_-'\chi_+')^2}
+ \frac{8}{s^2}(tt_1+uu_1-s^2-ss_1), \\ \nonumber
&& \Delta_{ss_1} = \frac{8}{s_1}\biggl(\frac{t}{\chi_-\chi_-'}
+ \frac{t_1}{\chi_+\chi_+'} - \frac{u}{\chi_-\chi_+'}
- \frac{u_1}{\chi_+\chi_-'}\biggr)  \\ \nonumber
&& \qquad + \frac{8}{ss_1}\biggl[
\frac{2(t_1-u)+u_1-t}{\chi_-'} + \frac{2(t-u_1)+u-t_1}{\chi_+'} \\ \nonumber
&& \qquad + \frac{u_1+t_1-s}{2\chi_-}\biggl(\frac{u}{\chi_+'}
- \frac{t}{\chi_-'}\biggr)
+ \frac{u+t-s}{2\chi_+}\biggl(\frac{u_1}{\chi_-'}
- \frac{t_1}{\chi_+'}\biggr)\biggr].
\end{eqnarray}
In the ultra--relativistic limit $(s\gg m_{\pi}^2)$ one has to drop
in the above formula the terms $\Delta_{ss}$, $\Delta_{s_1s_1}$,
$\Delta_{ss_1}$.

For numerical estimations it is useful to separate the most singular
part of the differential cross--section and to integrate it
analytically. We mean the contribution due to hard collinear
photon emission by electrons. We suggest the following procedure.
Let us define narrow cones surrounding the momenta of the initial
particles. Their opening angle is defined by an auxiliary
parameter $\theta_0$. The vertex is taken in the interaction point.
A photon emitted by the electron or the positron
inside the cones ($\widehat{\vecc{k}\vecc{p}}_- < \theta_0$ or
$\widehat{\vecc{k}\vecc{p}}_+ < \theta_0$) will be called as
a collinear one. On the parameter $\theta_0$ one has to impose
the restrictions
\begin{eqnarray}
1 \gg \theta_0 \gg \frac{m_e}{\varepsilon}\, .
\end{eqnarray}
Integrating inside the cones we drop all terms proportional to
$\theta_0^2$~\cite{quasi}. After simple calculations one comes to a formula,
where a factorization of a {\em shifted\/} Born cross--section can
be found. So, the process of collinear
photon emission is factorized with respect to the hard process
of the annihilation into pions. This is the manifestation of the known
factorization theorem~\cite{QCD}.
Indeed, we obtain the factorization
of large logarithm $L =\ln(s/m_e^2)$ with an accompaniment of several
non--leading terms. The dependence on the auxiliary parameter $\theta_0$
should cancel in the sum with the contribution of the integration outside
the cones. We will keep $\ln(\theta_0^2/4)$ terms and use them below as
a compensator.

The {\em shifted\/} cross--section $\dd\tilde{\sigma}(z_1,z_2)$ reads
\begin{eqnarray}
\dd\tilde{\sigma}(z_1,z_2)=\frac{\alpha^2}{4s}
\frac{(Y_1^2-m_{\pi}^2/\eps^2)^{3/2}}{z_1^2z_2^2}\;
\frac{(1-c^2)\dd\Omega|F_{\pi}(sz_1z_2)|^2}{z_1+z_2+(z_2-z_1)
(1-m^2_{\pi}/(\eps^2Y_1^2))^{-1/2}c}\, ,
\end{eqnarray}
where $z_1$ and $z_2$ are energy fractions of the {\em almost
real\/} electron and positron after the emission of collinear photons.
The energy fractions $Y_{1,2}$ of the final pions can be found
from the following kinematical relations:
\begin{eqnarray*}
&& y_{1,2}^2 = Y_{1,2}^2-\frac{4m_{\pi}^2}{s},\quad
z_1+z_2 = Y_1+Y_2,\qquad z_1-z_2=y_1c_- +y_2c_+, \\
&& y_1\sqrt{1-c_-^2} = y_2\sqrt{1-c_+^2},\quad c_-\equiv c, \quad
Y_{1,2}=\frac{q_{-,+}^0}{\eps}\, , \quad
c_+=\cos\widehat{\vecc{p}_-\vecc{q}}_+\, , \\
&& Y_1= - \frac{4m_{\pi}^2}{s}\;
\frac{(z_1-z_2)c}{2z_1z_2+[4z_1^2z_2^2-4(m_{\pi}^2/s)((z_1+z_2)^2
-(z_1-z_2)^2c^2)]^{1/2}}\\
&& \quad + \frac{2z_1z_2}{z_1+z_2-c(z_1-z_2)}\, .
\end{eqnarray*}

The leading contributions to cross--section, containing large
logarithm $L$, as may be recognized, combine into the kernel of
Altarelli--Parisi--Lipatov evolution equation:
\begin{eqnarray}
\dd\sigma &=& \int \dd z_1 \dd z_2 {\cal D}^{\gamma}(z_1)
{\cal D}^{\gamma}(z_2) \dd\tilde{\sigma}_0(z_1,z_2), \nonumber \\
{\cal D}^{\gamma}(z)&=&\delta(1-z)
+\frac{\alpha}{2\pi}(L -1) P^{(1)}(z)
+\left(\frac{\alpha}{2\pi}\right)^2\frac{(L -1)^2}{2!}P^{(2)}(z)
+ \dots\ ,
\\ \nonumber
P^{(1)}(z)&=&\lim\limits_{\Delta \to 0}\left(\delta(1-z)(2\ln\Delta
+ \frac{3}{2}) + \Theta(1-z-\Delta)\frac{1+z^2}{1-z}\right), \\ \nonumber
P^{(2)}(z)&=&\int\limits_{x}^{1}\frac{\dd t}{t}P^{(1)}(t)P^{(1)}
\left(\frac{z}{t}\right).
\end{eqnarray}
This formula is valid in the leading logarithmical approximation.
We will modify it by including nonleading contributions and using the
smoothed exponentiated representation for structure functions~\cite{strfun}:
\begin{eqnarray}
{\cal D}(z,s)&=&{\cal D}^{\gamma}(z,s)+{\cal D}^{e^+e^-}(z,s), \\ \nonumber
{\cal D}^{\gamma}(z,s)&=&\frac{1}{2}b\biggl(1-z\biggr)^{\frac{b}{2}-1}
\biggl[1+\frac{3}{8}b + \frac{b^2}{16}\biggl(\frac{9}{8}
- \frac{\pi^2}{3}\biggr)\biggr] \\ \nonumber
&-& \frac{1}{4}b(1+z)
+ \frac{1}{32}b^2\biggl(4(1+z)\ln\frac{1}{1-z}
+ \frac{1+3z^2}{1-z}\ln\frac{1}{z} - 5 - z \biggr), \\ \nonumber
{\cal D}^{e^+e^-}(z,s)&=&\frac{1}{2}b\biggl(1-z\biggr)^{\frac{b}{2}-1}
\biggl[ - \frac{b^2}{288}(2L - 15) \biggr] \\ \nonumber
&+& \left(\frac{\alpha}{\pi}\right)^2\biggl[
\frac{1}{12(1-z)}\biggl(1-z-\frac{2m_e}{\eps}\biggr)^{\frac{b}{2}}
\biggl(\ln\frac{s(1-z)^2}{m_e^2}-\frac{5}{3}\biggr)^2 \\ \nonumber
&\times& \biggl(1+z^2+\frac{b}{6}\biggl(\ln\frac{s(1-z)^2}{m_e^2}
-\frac{5}{3}\biggr)\biggr) + \frac{1}{4}L^2\biggl(\frac{2}{3}\;
\frac{1-z^3}{z} + \frac{1}{2}(1-z) \\ \nonumber
&+& (1+z)\ln z\biggr) \biggr]
\Theta(1-z-\frac{2m_e}{\eps}), \qquad
b = \frac{2\alpha}{\pi}(L-1).
\end{eqnarray}
In comparison with the corresponding formula in Ref.~\cite{strfun}
we shifted the terms, arising due to virtual $e^+e^-$ pair production corrections, form ${\cal D}^{\gamma}$
into ${\cal D}^{e^+e^-}$.

The final expression for the corrected cross--section reads as follows:
\begin{eqnarray}
\dd\sigma &=& \!\int\limits_{z_{\mathrm{min}}}^{1}\!\! \dd z_1
\!\int\limits_{z_{\mathrm{min}}}^{1}\!\! \dd z_2
{\cal D}(z_1,s){\cal D}(z_2,s) \dd\tilde{\sigma}(z_1,z_2)
\biggl(1+\frac{2\alpha}{\pi}( k(c,sz_1z_2) \nonumber \\ \nonumber
&+& b(sz_1z_2) + a)\biggr)\Theta_{\mathrm{cut}}(z_1,z_2)
+ \Biggl[ \frac{\alpha^3}{2\pi^2s^2}\!\!\!\!\!
\int\limits_{\stackrel{k^0>\Delta\eps}
{\widehat{\vecc{k}\vecc{p}}_{\pm}>\theta_0}}\!\!\!\!\!
R_1\bigg|_{m_e^2=0} \!\!\!
\dd\Gamma\;\Theta_{\mathrm{cut}}^{(5)} \\ \nonumber
&+& \frac{\alpha}{\pi}\int\limits_{\Delta\eps/\eps}^{1}
\frac{\dd x}{x}\left(1-x+\frac{x^2}{2}\right)\ln\frac{\theta_0^2}{4}
\biggl(\dd\tilde{\sigma}(1-x,1)\Theta_{\mathrm{cut}}(1-x,1) \\ \nonumber
&+& \dd\tilde{\sigma}(1,1-x)\Theta_{\mathrm{cut}}(1,1-x)\biggr)
\Biggr]
+ \Biggl[ \frac{\alpha^3}{2\pi^2s^2}
\int\limits_{k^0>\Delta\eps}\!\! R_2
\dd\Gamma\;\Theta_{\mathrm{cut}}^{(5)}
\\ \nonumber
&+& \frac{\alpha}{\pi}
2\ln\frac{\Delta\eps}{\eps}\dd\tilde{\sigma}(1,1)
\biggl(2\ln\frac{1-\beta c}{1+\beta c}
+ \frac{1+\beta^2}{2\beta}\ln\frac{1+\beta}{1-\beta} - 1 \biggr)
\Biggr]
\\ \label{eepipi}
&+& \frac{\alpha^3}{2\pi^2s^2}\int R_{3}\dd\Gamma
\;\Theta_{\mathrm{cut}}^{(5)}, \qquad
z_{\mathrm{min}} = \frac{2m_{\pi}}{2\eps-m_{\pi}}\, .
\end{eqnarray}
Quantities $a$, $b$ and $k$ are defined in Eqs.(\ref{evebsv},\ref{oddsv}).
Schematically this expression can be written as
\begin{eqnarray} \label{schem}
\dd\sigma = \dd\sigma^{(1)} + [\dd\sigma^{(2)}+C^{(2)}]
+ [\dd\sigma^{(3)}+C^{(3)}] + \dd\sigma^{(4)},
\end{eqnarray}
where quantities $C^{(2,3)}$ denote compensators; and terms
$\dd\sigma^{(2,3,4)}$ denote integrals of $R_{1,2,3}$,
respectively.
Starting with Eq.(\ref{rss}) we can replace in the above expression
quantities $R_i$ in the following way:
\begin{eqnarray}
R_1\bigg|_{m_e^2=0} \rightarrow \frac{s}{8} R_{s_1s_1}\bigg|_{m_e^2=0},
\qquad R_2 \rightarrow \frac{s}{8} ( R_{ss} + R_{ss_1} ), \qquad
R_3 \rightarrow 0.
\end{eqnarray}
Experimental conditions of the final particle
detection are encoded by the $\Theta$--functions
$\Theta_{\mathrm{cut}}(z_1,z_2)$ (for the two--particle final state
kinematics) and $\Theta_{\mathrm{cut}}^{(5)}$
(for the three--particle one).
They can be imposed explicitly by
introducing the restriction of the following kind:
\begin{eqnarray}
\Theta_{\mathrm{cut}}=\Theta(Y_1-y_{\mathrm{th}})\Theta(Y_2-y_{\mathrm{th}})
\Theta(\sin\theta_+ - \sin\Psi_0)\Theta(\sin\theta_- - \sin\Psi_0),
\end{eqnarray}
where $y_{\mathrm{th}}\eps=\eps_{\mathrm{th}}$ is the threshold
of the detectors, angle
$\Psi_0$ determines the {\em dead\/} cones around beam
axes unattainable for detection. Angles $\theta_{\pm}$
define the polar angles of the pions. More detailed cuts can be
implemented in a Monte Carlo program~\cite{MC}, using the
formulae given above.

There is a peculiar feature in the spectrum of hard photons.
Namely in the end of the spectrum the differential cross--section
is proportional to the factor
\begin{eqnarray}
I(s_1) = \frac{1}{s_1} \left(1-\frac{4m_{\mu}^2}{s_1}\right)^{3/2}\, ,
\end{eqnarray}
which defines a certain peak.
It comes from the Feynman diagrams describing the emission by
the initial particles~\cite{bkf}.

\section{Kaon Pair Production Near Threshold}

In the case of $K_LK_S$ meson pair production the differential
cross--section in the Born approximation reads
\begin{equation}
\frac{\dd\sigma_{0}(s)}{\dd\Omega_L}=\frac{\alpha^2\beta_{K}^3}{4s}
\sin^2\theta\; |F_{LS}(s)|^2.
\end{equation}
Here, as well as in the case of pions production, we suggest that the
form factor $F_{LS}$ includes also the vacuum polarization operator
of the virtual photon.
Quantity $\beta_{K}=\sqrt{1-4m_K^2/s}$ is the $K$--meson
velocity in the centre--of--mass frame,
and $\theta$ is the angle between the directions of motion of
the long living kaon and the initial electron.

The corrected cross--sections has the form:
\begin{equation}
\frac{\dd\sigma^{e^+e^-\to K_LK_S}(s)}{\dd\Omega_L}=
\int\limits_{0}^{\Delta}\dd x\;
\frac{\dd\sigma_0^{e^+e^-\to K_LK_S}(s(1-x))}{\dd\Omega_L}
F(x,s),
\end{equation}
where (see Ref.~\cite{strfun})
\begin{eqnarray}
F(x,s) &=& bx^{b-1}\biggl[ 1 + \frac{3}{4}b
+ \frac{\alpha}{\pi}\left(\frac{\pi^2}{3}-\frac{1}{2}\right)
- \frac{b^2}{24}\left(\frac{1}{3}L  - 2\pi^2 - \frac{37}{4}\right)
\biggr] \\ \nonumber
&-& b\left(1-\frac{x}{2}\right)
+ \frac{1}{8}b^2\biggl[4(2-x)\ln\frac{1}{x}
+ \frac{1}{x}(1+3(1-x)^2)\ln\frac{1}{1-x} - 6 + x\biggr] \\ \nonumber
&+& \left(\frac{\alpha}{\pi}\right)^2\biggl\{\frac{1}{6x}
\left(x-\frac{2m_e}{\eps}\right)^{b}\biggl[ (2-2x+x^2)
\left(\ln\frac{sx^2}{m_e^2}-\frac{5}{3}\right)^2
+ \frac{b}{3}\left(\ln\frac{sx^2}{m_e^2}-\frac{5}{3}\right)^3\biggr]
\\ \nonumber
&+& \frac{1}{2}L^2\biggl[\frac{2}{3}\;\frac{1-(1-x)^3}{1-x}
+ (2-x)\ln(1-x) + \frac{x}{2}\biggr]\biggr\} \Theta(x-\frac{2m_e}{\eps}).
\end{eqnarray}
We omitted a small contribution (proportional to
$\alpha(m_{\phi}-2m_K)/m_{\phi}$)
from photon emission by the final particles. Note that at higher
energies this effect is extremely interesting: it may shed
light on the neutral kaons polarizability problem.

In the case of $K^+K^-$ mesons production the Coulomb final state interaction
is to be taken into account:
\begin{eqnarray} \label{bornk}
\frac{\dd\sigma_{0}(s)}{\dd\Omega_-}&=&\frac{\alpha^2\beta_{K}^3}{4s}
\sin^2\theta |F_{K}(s)|^2\frac{Z}{1-\exp(-Z)}, \\ \nonumber
Z&=&\frac{2\pi\alpha}{v}, \qquad
v = 2\sqrt{\frac{s-4m_K^2}{s}}\,\left(1+\frac{s-4m_K^2}{s}\right)^{-1},
\end{eqnarray}
where $v$ is the relative velocity of
kaons~\cite{nuovo}. When $s=m_{\phi}^2$ we have $v\approx 0.5$.
Because we consider the energy range
close to $\phi$ mass one may choose the maximal energy of the soft
photon as
\begin{equation}
\omega\leq\Delta E=m_{\phi}-2m_K\ll m_K,\quad
\Delta\equiv\frac{\Delta E}{m_K}\approx\frac{1}{25}.
\end{equation}
If required, more precise formulae
for charge--even and charge--odd parts of cross--section
may be obtained from the ones for charged pions production process with the
replacement $\beta\to\beta_{K}$.

\section{Conclusions}

Thus, we presented differential
cross--sections to be integrated in concrete
experimental conditions. The formulae are good as for semi--analytical
integration, as well as for the creation of a Monte Carlo generator~\cite{MC}.
The idea of our approach was to separate the contributions
due to $2\to 2$ like processes and $2\to 3$ like ones.
The compensating terms allow us to eliminate the dependence
on auxiliary parameters in both contributions separately.
In the same approach we had considered in paper~\cite{lqed}
the processes of large--angle Bhabha scattering and
electron--positron annihilation into muons and photons.

Note that all presented formulae are valid only for large angle
processes. Indeed, in the region of very small angles
$\theta \approx m_e/\eps$ of final particles with respect
to the beam directions
there are contributions of double logarithmic approximation~\cite{quasi}.
These small angle regions give the main part of the total
cross--section. We suppose that this kinematics is rejected
by experimental cuts.

Numerical computations were done for the case of pion production.
In Fig.1 we give the differential cross--section in
the Born approximation, in the leading logarithmic approximation
(with the ${\cal K}$--factor) according to the first term of
Eq.(\ref{eepipi}), and as the total corrected cross--section.
We used the following set of parameters: the energy threshold for
pion registration $\Delta_1=y_{\mathrm{th}}=0.5$, the detector angular
acceptance $10^{\circ} < \theta_{\pm} < 170^{\circ}$, the beam c.m.s.
energy $\eps=0.51$~GeV. The second Figure shows the compensation
of auxiliary parameters. Value of $\sigma^{(2)}$ is shown by
the solid line. It has to be summed with $C^{(2)}$ which is drown
by circles. Value of $\sigma^{(3)}$ is shown by the dashed line.
It has to be summed with $C^{(3)}$ drown by crosses.
The parameters for Fig.2 are: $\Delta_1=0.5$, $\Delta=0.01$,
$\theta_0=0.01$, other parameters as in Fig.1.  The corrected
differential cross--section for charged pion production is shown
in Fig.3. The beam energy and angular acceptance are the same as
in Fig.1.
For these numerical illustrations we did not take into account
the pion form factor.

The precision of our results is defined by the omitted contributions.
As concerns pure QED terms, the corresponding uncertainty is
defined by unknown coefficients before the following terms:
\begin{eqnarray}
&& \left(\frac{\alpha}{\pi}\right)^2L \approx  10^{-4},\quad
\left(\frac{\alpha}{\pi}\right)^2  \approx  10^{-5},\quad
\frac{\alpha}{\pi}\,\frac{m_e^2}{s}L^2 \approx 10^{-7},
\\ && \frac{\alpha}{\pi}\theta_0^2\ln(\frac{4}{\theta_0^2})
\approx 10^{-5}, \qquad
\frac{\alpha}{\pi}\left(\frac{m_e}{\eps\theta_0}\right)^2\, .
\end{eqnarray}
We estimate the unknown coefficient\footnote{For small--angle
Bhabha scattering we had performed the complete calculations
of next--to--leading logarithmic terms~\cite{long}.
The corresponding coefficients were found to be about unity.}
and derive the theoretical uncertainty of our calculations
for these three processes to be $0.2 \%$.
Some additional uncertainty is due to precision of
the form factor determination in an experiment.
There are also some model--dependent contributions
due to different hadronic intermediate states,
which are far beyond the scope of this publication.

%\ack
\subsection*{Acknowledgement}
The authors are grateful to A.~Sher, S.~Panov for their close
interest in the initial stage, to V.S.~Fadin and
L.N.~Lipatov for critical comments and to O.O.~Voskresenskaya
for help. This work was  partially supported by INTAS grant
1867--93 and by RFBR grant $N^{\underline{\circ}}$~96--02--17512.
One of us (A.B.A.) is thankful to the INTAS foundation
for the financial support via the International Centre
for Fundamental Physics in Moscow.

\section*{Appendix}

Pion and kaon form factors are supposed to have the form
$F_{i}(s)=(1-\Pi(s))^{-1}{\cal F}_{i}(s)$,
where the factors
${\cal F}_{\pi,LS,K}(s)$ (unknown theoretically) take into account the
strong interactions between hadrons in the final state.
We put here the expressions for leptonic and hadronic contributions
into vacuum polarization operator $\Pi(s)$:
\begin{eqnarray}
\Pi(s) &=& \Pi_l(s) + \Pi_{h}(s), \\ \nonumber
\Pi_l(s) &=& \frac{\alpha}{\pi}\Pi_1(s)
+ \left(\frac{\alpha}{\pi}\right)^2\Pi_2(s)
+ \left(\frac{\alpha}{\pi}\right)^3\Pi_3(s) + \dots \\ \nonumber
\Pi_h(s) &=& \frac{s}{4\pi\alpha}\Biggl[
\mathrm{PV}\int\limits_{4m_{\pi}^2}^{\infty}
\frac{\sigma^{e^+e^-\to\mathrm{hadrons}}(s')}{s'-s}\dd s'
- \mathrm{i}\pi\sigma^{e^+e^-\to\mathrm{hadrons}}(s)\Biggr].
\end{eqnarray}
The first order leptonic contribution is well known~[1]:
\begin{eqnarray}
\Pi_1(s) &=& \frac{1}{3}L - \frac{5}{9} + f(x_{\mu}) + f(x_{\tau})
- \mathrm{i}\pi\left[\frac{1}{3} + \phi(x_{\mu})\Theta(1-x_{\mu})
+ \phi(x_{\tau})\Theta(1-x_{\tau})\right],
\end{eqnarray}
where
\begin{eqnarray*}
f(x) &=& \left\{\begin{array}{l}
-\frac{5}{9}-\frac{x}{3}+\frac{1}{6}(2+x)\sqrt{1-x}\ln\left(
\frac{1+\sqrt{1-x}}{1-\sqrt{1-x}}\right)\ \ \ \mathrm{for}\ \ x\leq 1, \\
-\frac{5}{9}-\frac{x}{3}+\frac{1}{6}(2+x)\sqrt{1-x}\;\mathrm{atan}\left(
\frac{1}{\sqrt{x-1}}\right)\ \ \ \mathrm{for}\ \  x > 1,\\
\end{array}\right. \\
\phi(x) &=& \frac{1}{6}(2+x)\sqrt{1-x},\qquad
x_{\mu,\tau} = \frac{4m_{\mu,\tau}^2}{s}\, .
\end{eqnarray*}
In the second order it is enough to take only the logarithmic term
from the electron contribution
\begin{eqnarray}
\Pi_2(s) = \frac{1}{4}(L-\mathrm{i}\pi) + \zeta(3) - \frac{5}{24}\, .
\end{eqnarray}

Here we present also a theoretical estimate for the contribution
to the vacuum polarization operator due to $\phi$ meson
in the energy region close to the resonance $\sqrt{s}\approx m_{\phi}
\approx 1020$~GeV (see also~\cite{fpi,phi1}).
In this region one can write the cross--section
of $e^+e^-$ annihilation into hadrons as follows:
\begin{eqnarray}
\sigma_{h}(s)=\frac{12\pi B_{ee} \Gamma^2_{\phi}}{(s-m^2_{\phi})^2
+m^2_{\phi}\Gamma^2_{\phi}}\, ,
\end{eqnarray}
where $B_{ee}$ is the branching ratio of the decay $\phi\to e^+e^-$,
quantity $\Gamma_{\phi}$ is the total width of the meson.
For the hadronic part of
vacuum polarization connected with $\phi$ meson we obtain:
\begin{eqnarray}
\Pi_h(s)=\Pi_{\phi}(s)+\Pi_{h'}(s),
\end{eqnarray}
where $\Pi_{h'}(s)$ includes other hadronic contributions~\cite{vacpol}.
The contribution $\Pi_{\phi}(s)$ is essential only in the region
$m_{\phi}-n\Gamma_{\phi}<\sqrt{s}<  m_{\phi}+n\Gamma_{\phi}$, $n\sim 1$.
It has the following form:
\begin{eqnarray}
\Pi_{\phi}(s)=\frac{3B_{ee}}{\alpha}\;\frac{ \frac{\Gamma_{\phi}}{m_{\phi}}
\left(\frac{s}{m_{\phi}^2}-1\right)}
{\left(\frac{\Gamma_{\phi}}{m_{\phi}}\right)^2 +
\left(\frac{s}{m_{\phi}^2}-1\right)^2}\, ,\qquad
B_{ee}\approx 3.09\cdot 10^{-4}.
\end{eqnarray}

%\end{document}

\newpage

\begin{figure}[t]
\begin{center}
\mbox{\epsfig{file=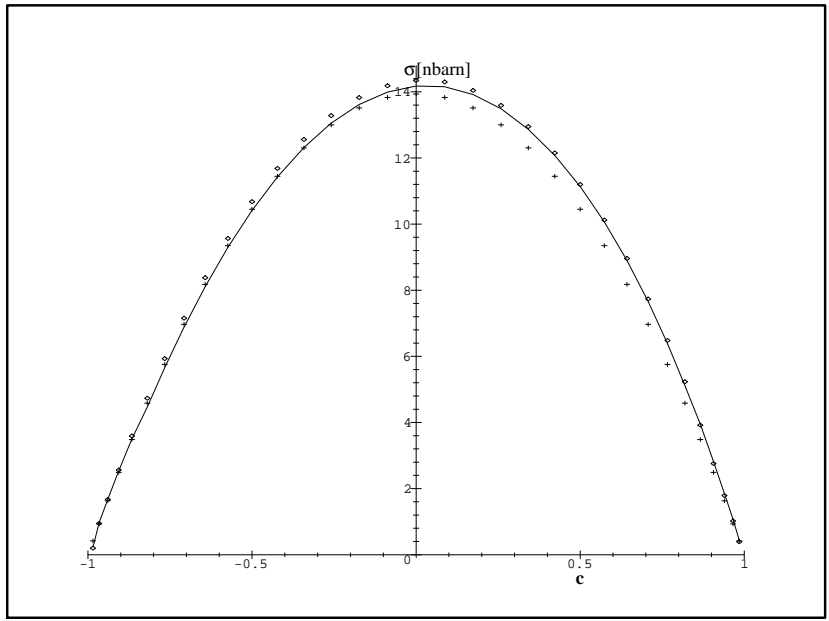,height=12cm,angle=270}}
\end{center}
\caption{The differential cross--section
$\dd\sigma^{e^+e^-\to\pi^+\pi^-}/\dd c$ [nbarn]
as a function of $c$.
The Born cross--section is given by crosses.
The cross--section in the leading logarithmic
approximation (with the ${\cal K}$--factor) is
shown by circles. The solid line is the resulting
corrected cross--section.
}
\label{Fig1}
\end{figure}

\newpage

\begin{figure}[t]
\begin{center}
\mbox{\epsfig{file=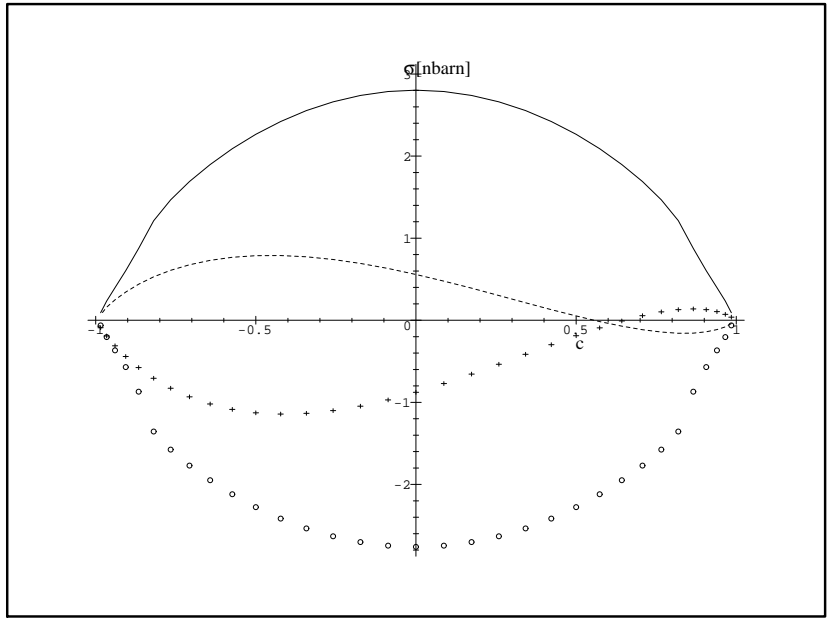,height=12cm,angle=270}}
\end{center}
\caption{
An illustration of the cancellation of the dependence on
$\Delta$ and $\theta_0$. The solid line is $\sigma^{(2)}$,
the dashed one is $\sigma^{(3)}$, compensator $C^{(2)}$ is
represented by circles and compensator $C^{(3)}$ is drown
by crosses.}
\label{Fig2}
\end{figure}

\newpage

\begin{figure}[t]
\begin{center}
\mbox{\epsfig{file=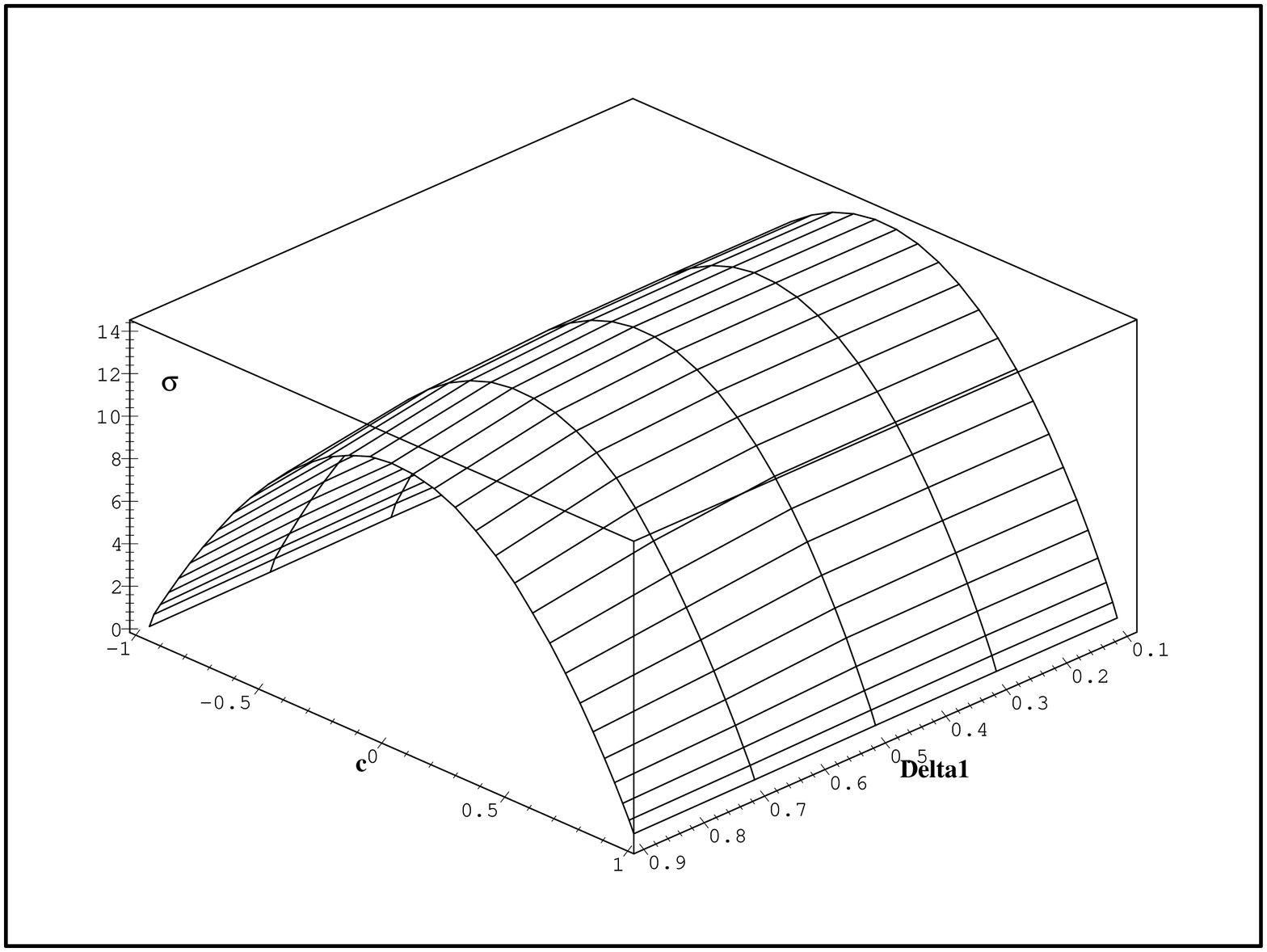,height=12cm,angle=270}}
\end{center}
\caption{
The value of the corrected cross--section
%(see Eq.(\ref{eepipi}))
$\dd\sigma^{e^+e^-\to\pi^+\pi^-}/\dd c$ [nbarn]
as a function of $c$ and $\Delta_1$.
}
\label{Fig3}
\end{figure}


\begin{thebibliography}{29}

\bibitem{vepp2m}
S.I.~Dolinsky et al.,
%{\em Summary of experiments with the neutral detector at
%     $e^+e^-$ storage ring VEPP--2M, \/}
Phys. Rep. {\bf 202} (1991) 99--170; \\
R.R. Akhmetshin et al.,
%{\em Measurement of $\phi$ meson parameters with
%     CMD--2 detector at VEPP--2M collider, \/}
Phys. Lett. {\bf B 364} (1995) 199.

\bibitem{dafne}
A.~Aloisio et al.,
%{\em KLOE, A General Purpose Detector for DA$\Phi$NE, \/}
preprint LNF-92/019 (IR); also in {\em The DA$\Phi$NE Physics
Handbook \/} Vol.2, 1993; \\
P.~Franzini,
%{\em The moun gyromagnetic ration and $R_H$ at DA$\Phi$NE$,\/}
in {\em The Second DA$\Phi$NE Physics Handbook,\/}
L.~Maiani, G.~Pancheri, N.~Paver (eds.), Vol.2, p.471, 1995.

\bibitem{bepc}
Hui--Ling Ni et al.,
Nucl. Instr. and Meth. {\bf A 336} (1993) 542.

\bibitem{lqed}
A.B.~Arbuzov, G.V.~Fedotovich, E.A.~Kuraev et al.,
% N.P.~Merenkov, V.D.~Rushai, L.~Trentadue,
% {\em Large angle QED processes at $e^+e^-$ colliders
%      at energies below 3 GeV,\/}
hep--ph/9702262.

\bibitem{fpi}
A.~Bramon, M.~Greco,
%{\em Electromagnetic Form Factors,\/}
in {\em The Second DA$\Phi$NE Physics Handbook,\/}
L.~Maiani, G.~Pancheri, N.~Paver (eds.), Vol.2, p.451, 1995. \\

\bibitem{prep}
E.A.~Kuraev, S.N.~Panov,
Preprint Budker INP 91--26, 1991.

\bibitem{int}
E.A.~Kuraev, preprint INP 80-155, Novosibirsk.

\bibitem{brown}
R.W.~Brown and K.O.~Mikaelian, Lett. Nouvo Cim. {\bf 10} (1974) 305.

\bibitem{quasi}
V.N.~Baier, V.S.~Fadin, V.A.~Khoze, Nucl. Phys. {\bf B 65} (1973) 381.

\bibitem{QCD}
G.~Sterman et al., Rev. Modern Phys. {\bf 67} (1995) 157.

\bibitem{strfun}
E.A.~Kuraev and V.S.~Fadin, Sov. J. Nucl. Phys. {\bf 41} (1985) 466.

\bibitem{MC}
V.A.~Astakhov, G.V.~Fedotovich et al.,
Monte Carlo generator for VEPP--2M, in progress.

\bibitem{bkf}
V.N.~Baier, V.M.~Katkov, V.S.~Fadin,
{\em Emission of the relativistic electrons,\/}
Moscow, Atomizdat, 1973.

\bibitem{nuovo}
A.B.~Arbuzov,
% {\em On a novel relativistic quasipotential equation
% for two scalar particles,\/} \\
Nuovo Cimento {\bf A 107} (1994) 1263.

\bibitem{vacpol}
S.~Eidelman, F.~Jegerlehner, Z. Phys. C 67 (1995) 585.

\bibitem{long}
A.B.~Arbuzov, V.S.~Fadin, E.A.~Kuraev, L.N.~Lipatov,
N.P.~Merenkov, L.~Trentadue,
%{\em Small--Angle Bhabha Scattering with a Per Mille Accuracy,\/}
preprint CERN--TH/95--313, to appear in Nucl. Phys. B.

\bibitem{phi1}
E. Drago, G. Venanzoni,
%{\em A BHABHA generator for DA$\Phi$NE including radiative
% corrections and $\phi$ resonance,\/}
KLOE MEMO $N^{\circ}$59/96.

\end{thebibliography}
\end{document}